\documentclass[twocolumn,runningheads]{svjour2}
\smartqed  
\usepackage{graphicx}
\journalname{Astrophysics and Space Science}
\begin{document}

\title{{\sl Chandra} Observations of Neutron Stars: An Overview.}

\author{M. C. Weisskopf \and M. Karovska \and  G. G. Pavlov \and 
         V. E. Zavlin \and T. Clarke
}


\institute{M. C. Weisskopf 
		\at VP60, MSFC, AL 23812 \\
		TEL.: +1 256-961-7798 \\
              \email{martin@smoker.msfc.nasa.gov}
           \and
           M. Karovska 
             \at SAO, 60 Garden St. Cambridge, MA 02138 \\
              Tel.: +1 617-495-7347 \\
              \email{karovska@head.cfa.harvard.edu}    
           \and
           G. G. Pavlov 
		\at PSU, 525 Davey Lab, University Park, PA 16802 \\
		TEL.: +1 814-865-9482 \\
              \email{pavlov@astro.psu.edu}
           \and
           V. E. Zavlin 
		\at NSSTC, 320 Sparkman Drive, Huntsville,AL 35805 \\
		TEL.: +1 256-961-7463 \\
              \email{slava.zavlin@nsstc.nasa.gov}
 \and
           T. Clarke 
\at Code 7213, NRL, 4555 Overlook Ave SW, Washington DC 20375 
\at Interferometrics Inc., 13454 Sunrise Valley Dr. \#240, Herndon, VA 20171 \\
		TEL.: +1 202-404-4297 \\
              \email{tracy.clarke@nrl.navy.mil}
}

\date{Received: date / Accepted: date}

\maketitle

\begin{abstract}
We present a brief review of {\sl Chandra} X-ray Observatory
observations of neutron stars.
The outstanding spatial and spectral resolution of this great observatory have allowed for observations of unprecedented clarity and accuracy. 
Many of these observations have provided new insights into neutron star physics. 
We present an admittedly biased and overly brief overview of these observations, highlighting some new discoveries made possible by the Observatory's unique capabilities.
We also include our analysis of recent multiwavelength observations of the putative
pulsar and its pulsar-wind nebula in the IC 443 SNR.


\end{abstract}

\section{Introduction}
\label{intro}
\begin{figure}
\centering
  \includegraphics[width=8.0cm]{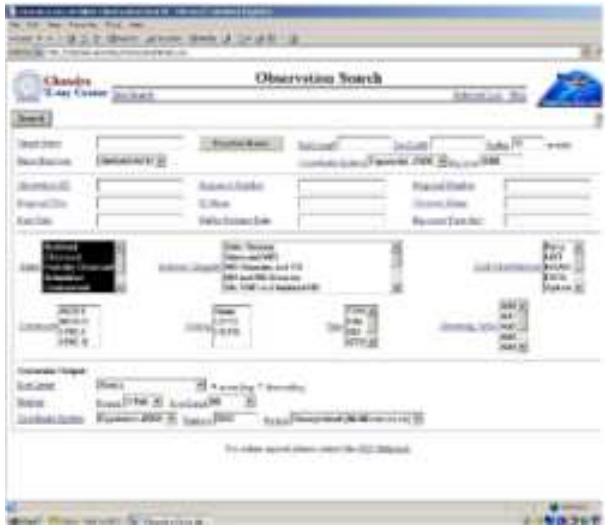}
\caption{{\sl Chandra} as perceived by an observer today.}
\label{f:fig1}
\end{figure}

\begin{figure}
\centering
  \includegraphics[width=8.0cm]{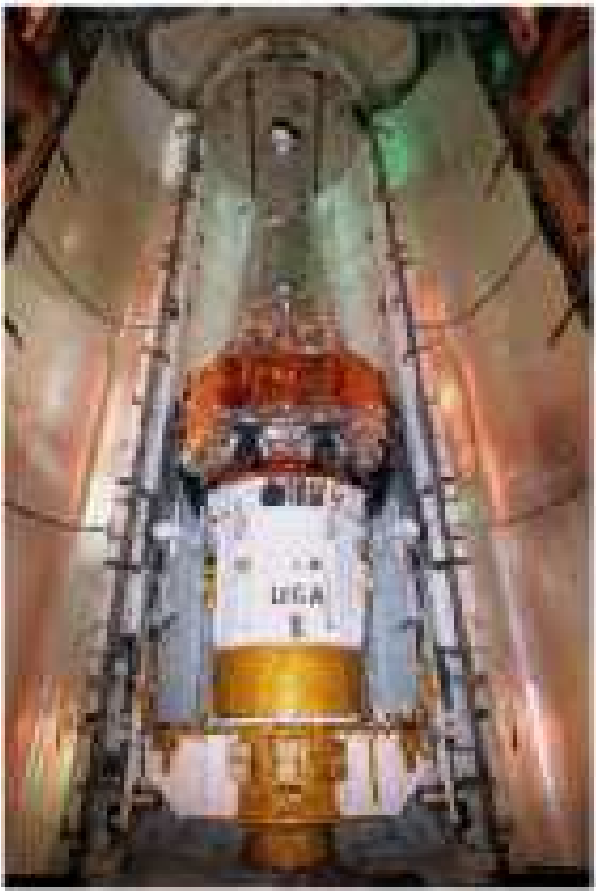}
\caption{{\sl Chandra} with IUS attached in Columbia's cargo bay.}
\label{f:fig2}
\end{figure}

For many users, the view of the {\sl Chandra} X-Ray observatory is as pictured in Figure~\ref{f:fig1}.
We therefore provide Figure~\ref{f:fig2} to remind us that the Observatory, with Inertial Upper Stage (IUS) attached, was the largest and heaviest payload ever launched by NASA's Space Transportation system.
The Observatory was launched on 1999 July 23 using the ill-fated orbiter Columbia. 
The IUS, a two-stage solid rocket booster, was subsequently fired and separated, sending {\sl Chandra} toward a high elliptical orbit.
On August 7, the fifth burn of {\sl Chandra}'s integral propulsion system placed it into a 63.5-hour (80,800-km semi-major axis) orbit. 
On August 12, the Observatory's forward door opened, exposing the focal plane to celestial X-rays.

The Observatory was designed for 3 years of operation with a goal of five years.
We are therefore extremely pleased that in July of this year the Observatory will have completed its seventh year of operation.
Certain difficulties (slow degradation of the thermal shielding and contamination buildup on the Advanced CCD Imaging Spectrometer [ACIS] instrument's cold filters) not withstanding, the Observatory continues to operate successfully.
The gas supply, used to point the Observatory, has a lifetime of much more than 15 years. 
The orbit will be stable for an even longer time period, and we anticipate many more years of usage.

\begin{figure}
\centering
  \includegraphics[width=8.0cm]{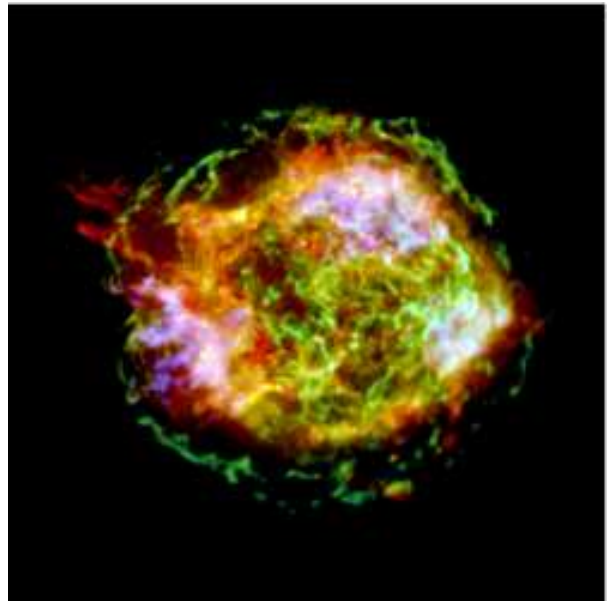}
\caption{{\sl Chandra} image of Cas A with the newly discovered point source at the center. Courtesy {\sl Chandra} X-Ray Center (CXC).
}
\label{f:casa}
\end{figure}

Even the first observations with {\sl Chandra} provided numerous startling insights concerning astrophysical systems in general, and neutron stars in particular. 
Two of the early images, those of the supernova remnant Cas A (Fig.\ \ref{f:casa}) and that of the Crab Nebula and its pulsar (Fig.\ \ref{f:crab}) have become two of the most spectacular and interesting observations made with {\sl Chandra}.

\section{The pulsar in the Crab Nebula.}
\label{s:crab}
The {\sl Chandra} X-Ray Observatory first observed the Crab Nebula and its pulsar during orbital calibration in 1999 (Weisskopf et al.\ 2000).
That image showed a striking richness of X-ray spatial structures: an X-ray inner ring within the X-ray torus; the suggestion of a hollow-tube structure for the torus; X-ray knots along the inner ring and (perhaps) along the inward extension of the X-ray jet.  
The {\sl Chandra} image also clearly resolved the X-ray torus 
and jet and counterjet which are all features that had been previously observed 
(Aschenbach \& Brinkmann 1975; Brinkmann, Aschenbach, \& Langmeier 1985; Hester et al.\ 1995; Greiveldinger \& Aschenbach 1999) but never with such clarity.
On slightly larger scales, the image also showed a sharply bounded notch (WSW of the Pulsar) into the X-ray nebular emission, earlier associated with the ``west bay'' of the Nebula (Hester et al.\ 1995).
Visible-light polarization maps of the Crab Nebula (Schmidt, Angel, \& Beaver 1979; Hickson \& van den Bergh 1990) demonstrate that the magnetic field is parallel to the boundary of this notch, thus serving to exclude the X-ray-emitting relativistic electrons from the west bay.

\begin{figure}
\centering
  \includegraphics[width=8.0cm]{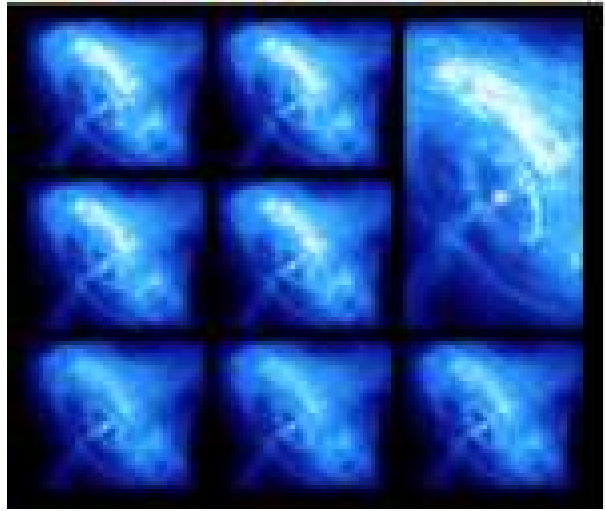}
\caption{Sequence of {\sl Chandra} ACIS observations of the Crab Nebula and pulsar. Courtesy CXC.
}
\label{f:crab}
\end{figure}

The most striking feature of the X-ray image is, of course, the inner elliptical ring, lying between the pulsar and the torus, 
which can be interpreted as a termination shock in the pulsar wind.
The existence of a termination shock had been predicted (Rees \& Gunn 1974; Kennel \& Coroniti 1984), but such a narrow ring, which can form
only in a wind confined in the equatorial plane, had never been expected. 
On the ring reside a few compact knots (Fig.\ \ref{f:crab}); one of them lying SE of the pulsar along the projected inward extension of the jet.
The surface brightness of this knot is too high to be simply explained as the superposition of the ring's and jet's surface brightnesses. 
Ultimately the nature of these knots needs to be probed by means of high-resolution spectroscopy.

Subsequently, Tennant et al.\ (2001) observed the system with {\sl Chandra} using the Low Energy Transmission Grating Spectrometer (LETGS).  
Time-resolved zeroth-order images were used to perform a most sensitive search for X-ray emission from the pulsar as a function of pulse phase, including pulse phases that had been traditionally referred to as ``unpulsed''; thus, as in the visible (Golden, Shearer, \& Beskin 2000; Peterson et al.\ 1978), the pulsar emits X-rays at {\em all} pulse phases.

Tennant et al. (2001) also confirmed prior observations (Pravdo, Angelini, \& Harding 1997; Massaro et al.\ 2000) which showed that the pulsar's power-law spectral index varied with pulse phase and also extended the measurements into the pulse minima.  
Finally, assuming that all of the flux from the pulsar at pulse minimum is attributable to thermal emission, the authors used these data to set a new upper limit to the blackbody temperature.  
As a representative case, they took the apparent angular radius of the neutron star 
$\theta_{\infty} = 2.1\times 10^{-16}\, {\rm rad}$ (corresponding to $R_{\infty} = 13\, {\rm km}$ at $D = 2\, {\rm kpc}$) and $N_{H} = 3\times10^{21}\, {\rm cm}^{-2}\!$.
With these parameters, the blackbody temperature
and luminosity that would account for all the flux observed at the pulse minimum were $T_{\infty} = 2.12\, {\rm MK} \ (kT_{\infty}= 183\, {\rm eV})\!$ and $L_{\infty} \approx 2.4 \times 10^{34}\, {\rm erg\ s}^{-1}\!$), which bounds the actual temperature and thermal luminosity of the neutron star.
Subsequent {\sl Chandra} LETGS observations and analyses of the spectrum as a function of pulse phase (Weisskopf et al.\ 2004) slightly improved this upper limit to  $T_{\infty} < 1.76$ and 2.01 MK at $2\sigma$ and $3\sigma$ confidence levels.\footnote{These upper limits appear weaker than previous {\sl ROSAT}-established upper limits set by Becker and Aschenbach (1995). The {\sl ROSAT} limits were, however, too ``optimistic'' as discussed in Tennant et al.\ (2001).}

Weisskopf et al.\ (2004) also performed a detailed analysis of the phase-averaged spectrum. 
They were able to study the interstellar X-ray extinction due primarily to photoelectric absorption and secondarily to scattering by dust grains in the direction of the Crab Nebula.  
They confirmed the findings of Willingale et al.\ (2001) that the line-of-sight to the Crab is under-abundant in oxygen.  
Using the abundances and cross sections from Wilms, Allen, \& McCray (2000), they found [O/H] = ($3.33\pm 0.25$)$\times 10^{-4}$.  
Spectral studies such as this, where the abundances are allowed to vary, are important as it is unlikely that standard abundances apply equally well to all lines of sight, especially those that intersect large quantities of SN debris (for more on this point see the discussion in Serafimovich et al.\ 2004).

In 2002, Hester et al.\ (2002) completed one phase of a set of coordinated observations of the Crab's pulsar wind nebula (PWN) using {\sl Chandra} (ACIS-S in sub-array mode) and the {\sl Hubble Space Telescope}.  
These spectacular observations revealed numerous dynamical features including wisps
moving outward from the inner equatorial X-ray ring with a speed of about 0.5 c.
These {\sl Chandra} observations are summarized in Figure~\ref{f:crab}.

Finally, {\sl Chandra} (and {\sl XMM-Newton}, see Willingale et al.\ 2001) has been used to study spectral variations as a function of position in the nebula.
Weisskopf et al.\ (2000) first presented the variation of a hardness ratio (the ratio of flux in two energy bands) as a function of position as seen with {\sl Chandra} using  $5''\times 5''$ pixels.
Mori et al.\ (2004) followed this work with studies of the variation of the power-law
spectral index as a function of position using $2.5'' \times 2.5''$ pixels.
It is worth mentioning that the pileup effect impacted the data analysis and required corrections.  
These corrections were not accurate when bright spatial structure was present within an analysis pixel --- dealing with that particular situation was noted by Mori et al. to be beyond the scope of their paper. 
One hopes this problem will be addressed by some enterprising expert in pileup in the future since {\sl Chandra} is providing one with unique information as to these small features.

\section{The Vela pulsar}
\label{ss:vela}
\begin{figure}
\centering
  \includegraphics[width=8.0cm]{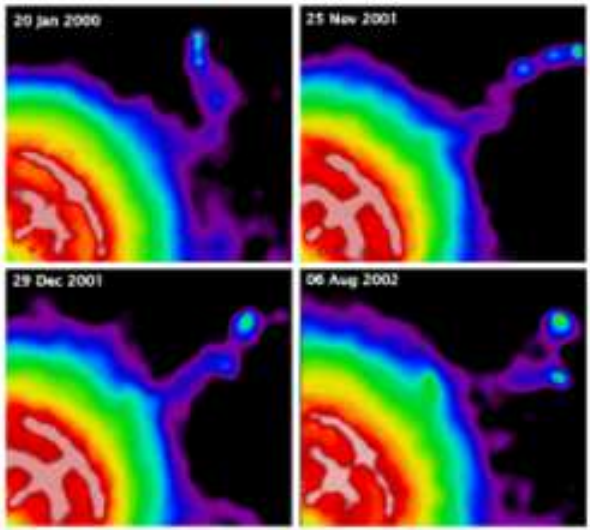}
\caption{Sequence of {\sl Chandra} ACIS observations of the Vela pulsar and its jet
(provided by O.\ Kargaltsev).
}
\label{f:vela}
\end{figure}

{\sl Chandra} observations of the 89-ms period Vela pulsar and its surroundings (Helfand, Gotthelf, \& Halpern 2001; Pavlov et al.\ 2001ab, 2003) have been most revealing. 
In addition to showing the complex and time variable spatial structure of the region immediately surrounding the pulsar itself --- a structure that includes two sets of arcs, a jet in the direction of the pulsar's proper motion and a counterjet --- the {\sl Chandra} images taken by Pavlov et al.\ (2003) also discovered that the continuation of the jet that extends to the NW is time-variable in both intensity and position on scales of days to weeks (Fig.\ \ref{f:vela}). 
Pavlov et al.\ (2003) found bright blobs in the outer jet, moving away from the pulsar, and inferred flow velocities of 0.3--0.7\,c.  
Finally, the apparent width of the outer jet appears to be approximately constant, despite large variations in appearance, indicating confinement.  
The analogy to a fire hose being held at its base is most appropriate.

\section{IC 443}
\label{s:ic443}
Historically, the first {\sl Chandra} observation of the IC 443 SNR (Olbert et al.\ 2001) was a publicity tour-de-force as the first three authors were high school students at the time.  
The remnant has been the subject of many studies in different wavelength bands as there is a large variety of shocked molecular gas present due to the interaction with surrounding molecular clouds (see references in Olbert et al.\ 2001, and Bykov, Bocchino, \& Pavlov 2005). 
IC 443 is also a candidate counterpart to the EGRET source 3EG J0617+2238.
The original {\sl Chandra} image clearly shows what appears to be a point source behind perhaps a bow shock and surrounded by a nebulosity that looks somewhat like a cometary tail. 
Olbert et al.\ (2001) also reported accompanying VLA observations which confirmed and complemented the X-ray spatial structure.
No pulsations were reported either from the X-ray or the flat spectrum radio observations.  
Subsequent observations with both {\sl Chandra} and {\sl XMM} (Bocchino \& Bykov 2001; Bykov, Bocchino, \& Pavlov 2005) have also not detected pulsations making this a candidate radio-quiet neutron star with a PWN.

More recently, an additional {\sl Chandra} observation was made by Gaensler et al.\ (2006; hereafter G2006).
Based on this observation with better angular resolution (the target was much closer to the aimpoint), G2006 confirmed the previously reported X-ray structure and discussed additional morphological details of the presumed bow-shock PWN, which they described as being a ``tongue'' of bright emission close to the neutron star, enveloped by a larger, fainter ``tail'' (a somewhat weird anatomical configuration!). 
G2006 also (reasonably) assumed that the brightest (point-like) enhancement in the X-ray image is the underlying rotation-powered pulsar (that has not been observed to pulse) that is creating the nebula.
These authors further commented upon small extensions near this object, which might be evidence for polar jets such as those seen in the Crab. 

We have been performing our own independent analysis of the {\sl Chandra} data, supplemented by observations at other wavelengths and find a number of intriguing and puzzling morphological features, most of which have been missed/overlooked by previous observers.

\begin{figure}
\centering
  \includegraphics[width=8.0cm]{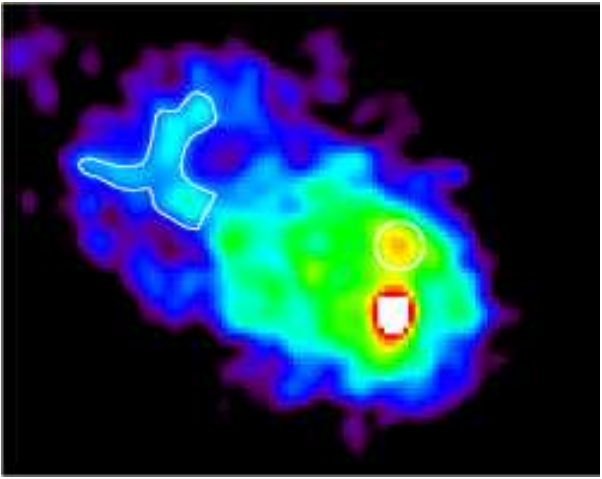}
\caption{$50''\times 40''$ image of the PWN core smoothed with a Gaussian of FWHM$=2''$. 
The white circle indicates the second point-like source $6''$ northward from the central and brightest source (presumably a pulsar), while the other white contour marks the ``hook''-like structure about $25''$ to the north-west from the pulsar.}
\label{f:ic443-core}
\end{figure}

{\bf Two Compact Sources in the X-ray PWN:} Our initial analysis of the early data showed the presence of not one, but two bright point-like enhancements in the X-ray data: the one reported by Olbert et al.\ (2001) and an apparently harder source $6''$ north of the putative pulsar.
Our conjecture was confirmed using the data from the longer more recent observation as shown in Figure~\ref{f:ic443-core}. 
It is not unreasonable to continue the assumption that the brighter enhancement to the south is the emission of the source powering the PWN, but it is worth investigating the northern feature more closely.
Spectral analysis of the approximately 300 counts in the southern enhancement by G2006 is certainly consistent with the pulsar interpretation, but by no means definitive given the small number of counts.
There are also not enough counts (between 20 and 50, depending on the background subtraction) from the northern source to determine spectral parameters and pursue the question as to the nature of this enhancement.
Is it associated with the PWN or it is an interloper? 
Is it connected with the southern source by some faint ``bridge'' (e.g., a receding pulsar jet)? 
Although unlikely, is it possibly the pulsar? 

\begin{figure}
\centering
  \includegraphics[width=8.0cm]{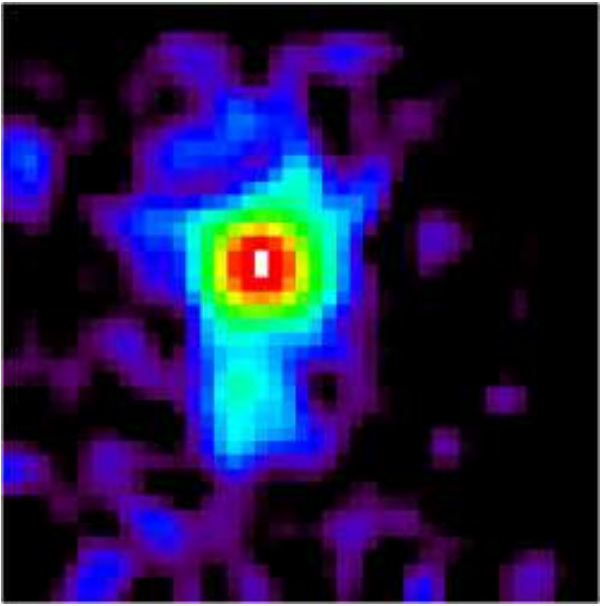}
\caption{ $8''\times 8''$ image of the southern source (pulsar) binned 
in 1/9 ACIS pixels and smoothed with a Gaussian of FWHM=$0.5''$. 
A $2''$-long structure extending southward from the pulsar is clearly visible.}
\label{f:ic443-jet}
\end{figure}

{\bf ``Jets'' in the Southern Source:} G2006 noticed short extensions towards north and south of the southern point-like source and suggested these extensions might be the jets emanating from the pulsar. 
Such an interpretation means that the jets (and hence the pulsar's spin axis) are not coaligned with the symmetry axis of the PWN, contrary to many other cases.
If so, this has important implications for our understanding of the ``natal kicks'' of neutron stars (the [mis]alignment constrains the kick timescale: 
if this timescale is larger than the pulsar period at birth, then the momentum imparted by the kick is rotationally averaged, resulting in alignment between velocity and spin vectors -- see, e.g., Romani 2005).
Our analysis of the same data convincingly reveals only the southern extended feature, with an estimated number of counts of $56\pm14$ (in agreement with G2006), while the existence of a northern ``jet'' is questionable (Fig.\ \ref{f:ic443-jet}).

\begin{figure}
\centering
  \includegraphics[width=8.0cm]{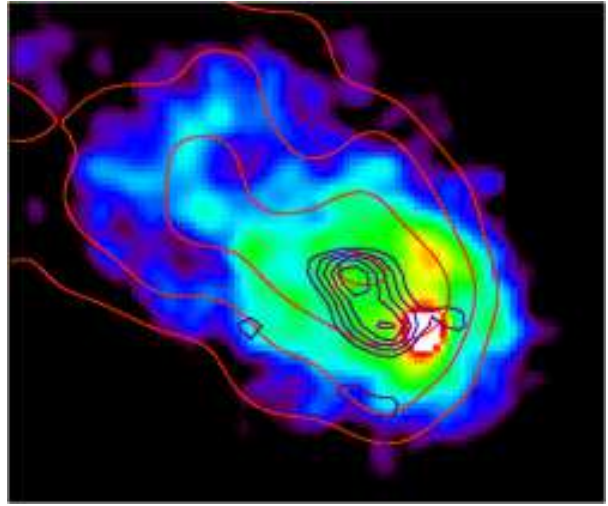}
\caption{Same as Fig.~\ref{f:ic443-core} with overlaid contours of radio brightness
measured at 8.6 (red) and 4.8 (blue) GHz.}
\label{f:ic443-radio}
\end{figure}

{\bf The radio/X-ray surface brightness puzzle:} In a ``classical'' PWN, such as the Crab and Vela, the radio emission tends to peak at the outskirts of the nebula due to synchrotron cooling (e.g., Kargaltsev \& Pavlov 2004).
Yet for IC443, the radio emission appears contained within the boundaries defined by the X-rays (Fig.\ \ref{f:ic443-radio}).
This might be explained by different energy distributions of relativistic wind electrons (e.g., more electrons with lower energies in the IC443 PWN than in the Crab and Vela).

\begin{figure}
\centering
  \includegraphics[width=8.0cm]{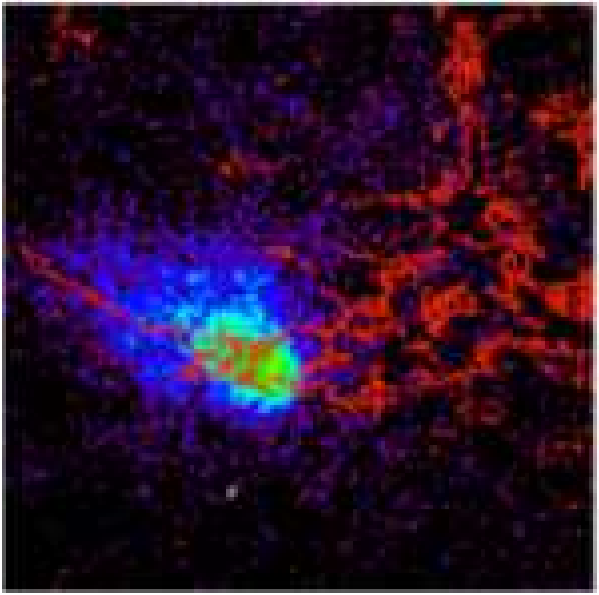}
\caption{X-ray (green and blue) and H$_{\alpha}$ (red) emission. 
The image size is approximately $5'\times 5'$.}
\label{f:ic443-halpha}
\end{figure}

{\bf The H$_{\alpha}$ filament and the PWN:} To our knowledge, we are the first to note the rather puzzling positional alignment between the X-ray PWN and the H$_{\alpha}$ filamentary structure shown in Figure~\ref{f:ic443-halpha}.
Optical observations (performed by J.\ Thorstensen and R.\ Fesen) show no hint of this structure at other wavelengths. 
Observationally, it is extremely unlikely that the major axis of the PWN and the H$_{\alpha}$ filamentary structure would be so aligned. 
Trying to understand a possible astrophysical reason is, however, a challenge. 
One might speculate that the H$_\alpha$ structure shows the streamlines of the SNR gas flowing round the PWN, excited by interaction with the pulsar wind and PWN radiation. 
The long narrow H$_\alpha$ tail behind (i.e. to the north-east in Figure~\ref{f:ic443-halpha}) the X-ray PWN would then mean that the excited gas is moving in a narrow channel, suggesting some collimation mechanism.
We have proposed to investigate this mystery further in a deeper {\sl Chandra} observation that would allow us to assess the correlation between the X-ray and H$_\alpha$ intensities at large distances behind the moving pulsar.

\section{B1509--58}
\begin{figure}
\centering
  \includegraphics[width=8.0cm]{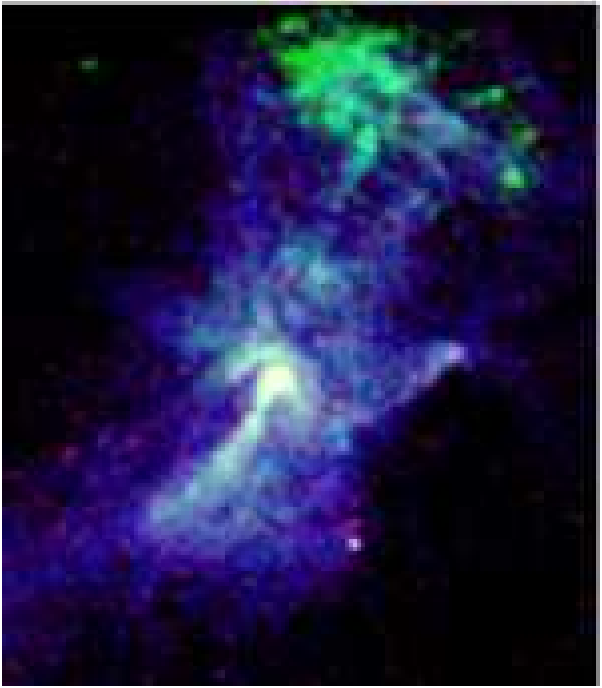}
\caption{{\sl Chandra} observation of the PWN powered by PSR B1509$-$58 in SNR 
320.4$-$1.2. Courtesy CXC.
}
\label{f:b1509}
\end{figure}

We complete our selection of {\sl Chandra} images of neutron stars and their pulsar-driven wind nebulae with the image (Fig.\ \ref{f:b1509}) of the 
PWN powered by a young, 150 ms pulsar B1509$-$58 in SNR 320.4$-$1.2 (Gaensler et al.\ 2002).
The pulsar is the bright source at the center of the nebula. 
A thin jet can be seen in the image to extend to the southeast. 
Just above the pulsar there is a small arc of X-ray emission, which seems to mark the location of the shock wave produced by the particles flowing away from the pulsar's equator. 
The cloud near the top of the image (known as RCW 89) may be due to high-temperature gas. 
This gas, possibly a remnant of the explosion associated with the creation of the pulsar, may have been heated by collisions with high-energy particles produced by the pulsar. 
Yatsu et al.\ (2005) provide a discussion of the interaction of the pulsar's jet with this material. 

\section{Finding Pulsars}
\label{s:findingpulsars}

\subsection{1E 1207.4$-$5209\label{ss:SNR296.5+10}}
Observations with {\sl Chandra} have contributed at least two significant new insights into the source 1E 1207.4$-$5209, initially discovered with the {\sl Einstein} Observatory (Hel-fand \& Becker 1984) and located $6'$ from the center of SNR PKS 1209$-$51/52 (aka SNR 296.5+10.0). 
The first observation of this source with {\sl Chandra}'s ACIS, taken 
in continuous clocking mode, resulted in the detection of a 424-ms period, 
which, of course, provided compelling evidence that the source is a neutron star
(Zavlin et al.\ 2000). 
Since the source appears to be radio-quiet (Mereghetti, Bignami, \& Caraveo 1996; Kaspi et al.\ 1996), it may be either an active pulsar beamed out of our line of sight (or perhaps a rotating radio transient [RRAT]!?) or a truly radio-quiet neutron star, where the X-ray pulsations are caused
perhaps by hot spots rotating in and out of our line of sight.
A second {\sl Chandra} observation allowed Pavlov et al.\ (2002a) to establish 
a preliminary estimate for the period derivative (hence, the spindown power,
characteristic age, and a dipole component of the magnetic field).

\begin{figure}
\centering
  \includegraphics[width=8.0cm]{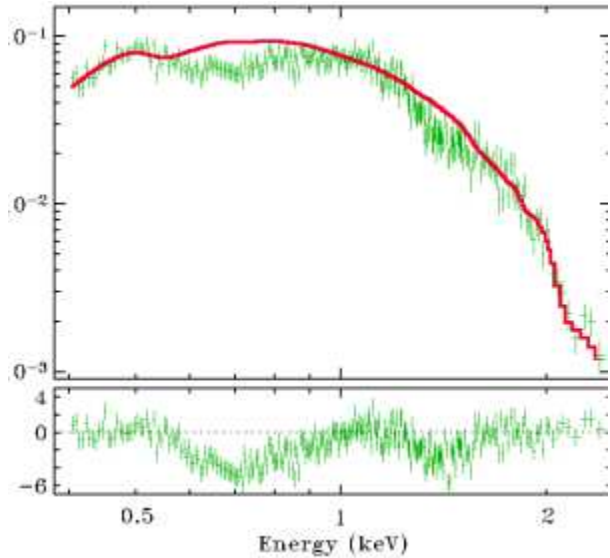}
\caption{The upper panel shows the data for 1E 1207.4$-$5209 and a featureless thermal spectrum. The lower panel displays the residuals to this spectrum and clearly indicates the presence of the two absorption features discussed in the text.
}
\label{f:1e1207spectrum}
\end{figure}

From the spectral analysis of the same two ACIS-S3 observations, Sanwal et al.\ (2002) found two significant absorption features centered at 0.7 and 1.4 keV with equivalent widths of about 0.1 keV (Fig.\ \ref{f:1e1207spectrum}).
These authors discussed several possible interpretations for the absorption, including cyclotron resonances and atomic features.  
They presented arguments favoring atomic transitions of once-ionized helium in the atmosphere of the neutron star assumed to be very strongly magnetized ($B\approx 10^{14}$\,G).
Other authors suggested different interpretations of the {\sl Chandra}-discovered features.
For example, Hailey and Mori (2002) argued that the absorption features were associated with He-like oxygen or neon in a field of $\approx 10^{12}$\,G.  
More recent {\sl XMM} observations (e.g., Mereghetti et al.\ 2002; Bignami et al.\ 2003; De Luca et al.\ 2004) not only confirmed the {\sl Chandra}-detected absorption features at 0.7 and 1.4 keV, but also seemed to have detected an additional feature at 2.1 keV and evidence for a fourth one, at 2.8
keV.  
Taking all these latter data into account supports an explanation involving the fundamental and two, possibly three, harmonics of the electron cyclotron absorption in a field of order $10^{11}$\,G. 
However, the two additional spectral features in the {\sl XMM-Newton} data have not been universally accepted. 
Mori, Chonko, \& Hailey (2005) have cast doubt as to the reality of the spectral features at 2.1 and 2.8 keV.  
Their arguments seem compelling, and it is thus unfortunate that the {\sl Chandra} response is insufficient to weigh in on this question without expending significant amounts of observing time.

Zavlin, Pavlov, \& Sanwal (2004) have continued to observe this target using both
{\sl Chandra} and {\sl XMM-Newton}.
They have detected significant non-monotonous variations in the spin period which they interpreted in light of three hypotheses: a glitching pulsar; variations in an accretion rate from a fallback disk; and variations produced by 
orbital motion in a wide binary (see also the presentation by Woods, Zavlin, \& Pavlov at this conference for new timing results).

We conclude this section by noting that the sequence of {\sl Chandra} observations have provided important discoveries --- the detection of the pulse period and firm detection of two absorption features.  
An important and unanswered question is what are the limits as to the presence of such spectral features for the other neutron stars in SNRs. 
A systematic comparison, if not already in progress, should be performed.

Finally we note that 1E 1207.4$-$5209 is a source that, in some critical respects, is similar to the central compact object (CCO) in the Cas~A SNR (Pavlov et al.\ 2000): 
it is in a SNR, it is radio-quiet, and shows a thermal-like X-ray spectrum (albeit with somewhat lower temperature); however it pulses.  
Thus, on the one hand, the source might give us confidence that all CCOs will ultimately be found to pulse.  
On the other hand, it may happen that the {\sl Chandra}-revealed characteristics
of this source will remain unique among the CCOs and will be used to separate it from that class of objects.
Time will tell.

\subsection{SNR 292.0+1.8}
\label{ss:snr292.0+1.8}
SNR 292.0+1.8 is, along with Cas A and Puppis A, an oxygen-rich supernova remnant.  Hughes et al.\ (2001) used {\sl Chandra} and detected a bright, spectrally hard, point source within an apparently extended region. 
This detection suggested the presence of a pulsar and its PWN.
Radio observations (Camilo et al.\ 2002) then found a 135-ms radio pulsar. 
The subsequent detection by Hughes et al.\ (2003) of X-ray pulses at the expected period secured the identification. 
The X-ray spectrum is modeled with a simple power-law, although, as with Vela (and many other sources!), the fit to the data is not unique. 
From the motions of oxygen-rich optical knots and the size of the remnant, Ghavamian, Hughes, \& Williams (2005) estimated a kinematic age for SNR 292.0+1.8 of about 3200 years, assuming a distance of 6 kpc. 
This value is in good agreement with the pulsar's spin-down age of 2900 years.

\subsection{3C58}
\label{ss:3c58}
The {\sl Chandra} observations of 3C58 (aka SNR 130.7+3.1) were first performed by Murray et al.\ (2002) using the High Resolution Camera (HRC). 
These data imaged the previously detected X-ray point source (Becker, Helfand, \& Szymkowiak 1982). 
The early {\sl Chandra} data also revealed an extended PWN and the presence of 66 ms pulsations from the central point source (J0205+6449). 
Deeper {\sl Chandra} observations, using ACIS-S3, by Slane et al.\ (2004) produced images ({Fig.\ \ref{f:3c58}) showing little similarity of this PWN with the Crab and Vela (in particular, the presence of numerous loop-like filaments), although one might speculate that we see a torus edge-on, projected in the north-south direction, and a one-sided jet westward of the pulsar.

\begin{figure}
\centering
  \includegraphics[width=8.0cm]{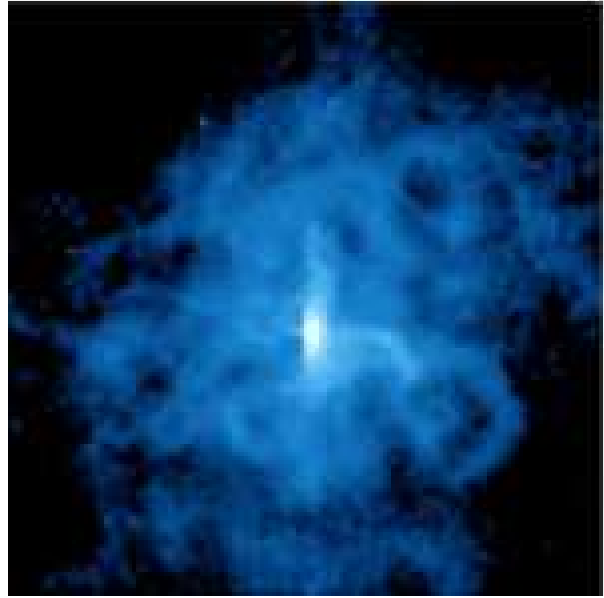}
\caption{{\sl Chandra} ACIS-S3 image of 3C58. Courtesy P.\ Slane.
}
\label{f:3c58}
\end{figure}

One aspect of the {\sl Chandra}-based research of 3C58 of special importance was the limits as to any thermal emission from the surface of the young neutron star, J0205+6449.
The search for thermal emission was first presented by Slane, Helfand, \& Murray (2002) and then refined by Slane et al.\ (2004) who found that 
the upper limits on surface temperature and thermal luminosity fall well below predictions of standard neutron star cooling. 
Yakovlev et al.\ (2002) discuss calculations of neutron star cooling in the context of J0205+6449 and concluded that the observations can be explained by the cooling of a superfluid neutron star where the direct Urca processes are forbidden.

Of course, neutron stars may be different, so that limits to the thermal components of 3C58 may not apply to all young neutron stars. 
In general, such analyses are not simple, requiring enhanced sensitivity for the detection of the putative thermal component often in the presence of a much stronger non-thermal flux from the magnetosphere of the pulsar, especially if one wants to measure the temperature --- as opposed to setting an upper limit. 
{\sl Chandra} is uniquely poised to provide the data for such studies due to its ability to maximally separate the pulsar from the surrounding nebulosity, yet often long observations are required.

\section{Not finding neutron stars} 
\label{sec:gnfns}
Because of its tremendous sensitivity facilitated by its superb optics that have such low scatter and therefore provide high contrast, {\sl Chandra} is ideally suited for searching for X-ray sources.

\subsection {$\gamma$-Cygni (SNR 78.2+2.1)}
\label{ss:gammacygni}
Becker et al.\ (2004) used {\sl Chandra} to search for the X-ray counterpart to the unidentified EGRET source 3EG J2020+4017.
These authors investigated 
the possibility (Brazier et al.\ 1996) that RX\, J2020.2+4026,  was the counterpart and concluded that it is associated with a K field star, excluding it from being counterpart of the bright $\gamma$-ray
source.

The observation also demonstrated the difficulties one often encounters in searching for compact objects associated with a SNR.
Thirty seven additional X-ray sources were detected in the field searched 
(which was only a fraction of the full size of the SNR). 
Radio observations reported by these authors, which covered the complete 99\% EGRET likelihood contour of 3EG J2020+4017 with a sensitivity limit of $L_{\rm 820\,MHz} = 0.09 \mbox{ mJy kpc}^2$, were unable to find a pulsar.  
The absence of radio pulsations suggests that if there is a pulsar operating in 
$\gamma$-Cygni, the pulsar's emission geometry is such that the radio beam does not intersect with the line of sight. 
Alternatively, the X-ray counterpart might be a CCO which does not produce significant amounts of radio emission although CCOs are not known to show $\gamma$-ray emission. 
(Or perhaps the source is the counterpart to an RRAT, and the radio observations were not performed at the appropriate time.)

Without high-precision X-ray spectra of each of the candidate X-ray sources, and detailed follow up in other wavelength bands, there is essentially no satisfactory way in which to eliminate most of the candidates from consideration. 
This is especially true for neutron stars as the the corresponding infrared-visible fluxes are very weak, making such observations especially difficult.
In such cases, the principal and important {\sl Chandra} contribution is to provide target lists with accurate positions as a basis for future studies.

\subsection{SNR 1987A}
\label{sec:1987a}
Since the launch of {\sl Chandra}, SN 1987A has been the focus of a series of repeated observations (see Park et al.\ 2005 and references therein).  
One goal of these observations is to detect the emergence of the X-ray flux from a newly born compact object.
Figure~\ref{f:snr-1987a} shows a sequence of such observations from 1999 through the middle of last year.
These images are a textbook demonstration of the development of the various shocks resulting from a supernova event.
The images also dramatically demonstrate that, thus far, no central compact object has yet appeared at X-ray wavelengths. 
Considering the spatial scale, these are observations that can only be performed with {\sl Chandra}, now and for the foreseeable future.

\begin{figure}
\centering
  \includegraphics[width=8.0cm]{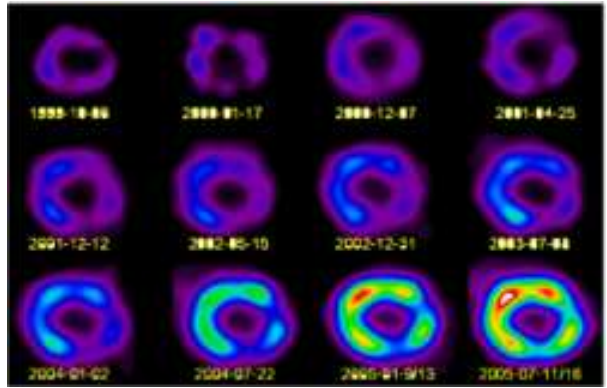}
\caption{Sequence of {\sl Chandra} observations of SNR 1987A (provided by S.\ Park).
}
\label{f:snr-1987a}
\end{figure}

\section{RX J1856.5--3754}
\label{sec:rxj1856}
RX J1856.5$-$3754 is the brightest of a number of ROSAT-detected objects, which are thermal emitters with blackbody temperatures in the range of 50--100 eV. 
These sources appear to be isolated neutron stars, and therefore RX J1856.5$-$3754 was a promising candidate for a long {\sl Chandra} observation in the hope that spectroscopy would provide a determination of parameters such as the radius, surface gravity, and gravitational redshift --- all of which would lead one to the holy grail of the equation of state. 
The featureless, 500-ksec spectrum that resulted (Fig.\ \ref{f:rxj1856-spectrum}, Burwitz et al.\ 2003) was therefore a surprise, although in retrospect it is worth noting that no features were detected in a much shorter (50-ksec; Burwitz et al. 2001) prior observation. 
(Ironically, spectral features were later found in at least 3 other, not so bright, sources of this class.)
The absence of line features needs an explanation, and has triggered new and interesting theoretical work. 
At one point the object was even claimed to be a quark star (Drake et al.\ 2002), but this appears to be no longer considered credible. 
Other suggestions include high ($\approx 10^{13}$ G) field strength (e.g., Tr\"{u}mper et al.\ 2003; see also Turolla, Zane, \& Drake 2004) neutron stars, 
and a slowly spinning ($P > 100,000$ sec) magnetar (Mori \& Ruderman 2003). 
Braje \& Romani (2002) and Pavlov, Zavlin, \& Sanwal (2002b) noted that several effects (e.g., fast rotation) may act in suppressing spectral features from an
atmosphere with heavy elements, but no period has been detected from this puzzling source.
Although it seems quite certain that RX J1856.5$-$3754 is a very compact object,
like a neutron star, its true nature still remains elusive.

\begin{figure}
\centering
  \includegraphics[width=8.0cm]{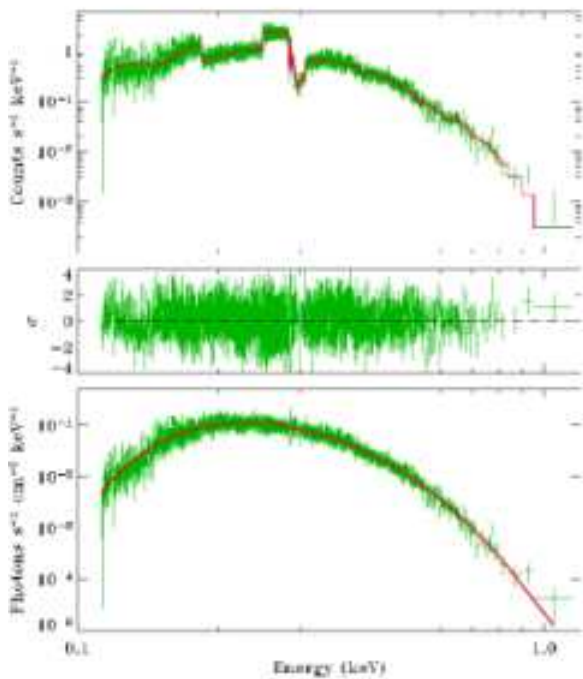}
\caption{{\sl Chandra} LETGS spectrum of RX J1856.5$-$3754 based on 500 ksec of observation. The upper panel shows the raw data, and the best-fit blackbody model. The middle panel shows the residuals to the best fit and the bottom panel the incident flux.
}
\label{f:rxj1856-spectrum}
\end{figure}

\begin{acknowledgements}
A number of the figures in this paper were provided by the {\sl Chandra} X-Ray Center (CXC) and are publicly available via http://chandra.harvard.edu.
MCW acknowledges support by the {\sl Chandra Project}.
MK is a member of the CXC, operated by the Smithsonian Astrophysical Observatory under contract NASA NAS8-39073.
The work of GGP was supported by NASA grant NAG5-10865.
VEZ is supported through a NASA Research Associateship Award. 
Basic research in radio astronomy performed by TC at NRL is supported by the Office of Naval Research.
\end{acknowledgements}

\end{document}